\documentclass[prl, twocolumn,superscriptaddress,amsmath,amssymb,showpacs,floatfix,preprintnumbers]{revtex4-2}
\usepackage{array}[=2016-10-06]

\usepackage{amsmath}
\usepackage{graphicx}
\usepackage{dcolumn}
\usepackage{xcolor}
\usepackage{physics}
\usepackage{xr}
\usepackage{siunitx}
\usepackage{tikz}
\usepackage{tabularx}
\usepackage{makecell}
\usepackage{multirow}
\usepackage{float}

\usepackage[
    colorlinks=true,
    linkcolor=blue,
    citecolor=blue,
    urlcolor=blue
]{hyperref}
\DeclareGraphicsExtensions{.png .jpg .pdf}

\begin{document}

\title{Symmetry-isolated magnetoelectric electro-optic effects in noncentrosymmetric metals}
\author{C. O. Ascencio}
\affiliation{School of Physics and Astronomy, University of Minnesota, Minneapolis, Minnesota 55455, USA}
\author{D. J. P. de Sousa}
\affiliation{Department of Electrical and Computer Engineering, University of Minnesota, Minneapolis, Minnesota 55455, USA}
\affiliation{Instituto de Física de São Carlos, Universidade de São Paulo, IFSC-USP, 13566-590, São Carlos, SP, Brazil }

\author{Seungjun Lee}
\affiliation{Department of Electrical and Computer Engineering, University of Minnesota, Minneapolis, Minnesota 55455, USA}
\affiliation{Department of Applied Physics, Kyung Hee University, Yongin 17104, Republic of Korea}
\author{Tony Low}\email{tlow@umn.edu}
\affiliation{School of Physics and Astronomy, University of Minnesota, Minneapolis, Minnesota 55455, USA}

\affiliation{Department of Electrical and Computer Engineering, University of Minnesota, Minneapolis, Minnesota 55455, USA}

\date{ \today }

\begin{abstract}
We classify the symmetry-constrained forms of the Berry curvature dipole $\mathbf{D}$, gyrotropic magnetic tensor $\mathbf{K}$, and magnetoelectric electro-optic (EO) tensor $\mathbf{G}$, which describe metallic optical and EO effects in time-reversal symmetric, noncentrosymmetric metals. We identify 11 space groups (SGs) in which $\mathbf{D}$ and $\mathbf{K}$ vanish by symmetry while $\mathbf{G}$ remains allowed, thereby providing a more direct route to observing the recently predicted magnetoelectric EO effects associated with $\mathbf{G}$. First-principles based calculations confirm that $\mathbf{D}$ and $\mathbf{K}$ vanish for representative materials, while $\mathbf{G}$ remains allowed and tunable via Fermi level shifting. We further show that the choices of SG and experimental configuration provide complementary paths for isolating $\mathbf{G}$-driven EO effects, including cases where $\mathbf{D}$ and $\mathbf{K}$ are also symmetry-allowed. In an oblique-incidence geometry, the $\mathbf{D}$-driven response produces a helicity-even absorption or gain correction, whereas the $\mathbf{G}$-driven response couples the $s$ and $p$ optical sectors and produces a bias-induced circular dichroism with a characteristic $\sin\theta\cos\theta$ angular dependence. This provides a direct experimental route for separating the $\mathbf{D}$- and $\mathbf{G}$-driven EO signatures in noncentrosymmetric metals.
\end{abstract}

\setcounter{secnumdepth}{3}
\maketitle
\section{Introduction}

Berry curvature and magnetic moment textures of Bloch electrons provide a route to optical responses and transport that are not captured by conventional Drude theory alone~\cite{Berry, AHE, GME1, GME2, Rappoport2023, Morgado2024, Fu, chiral_optical}. In noncentrosymmetric metals, these quantities can remain finite on the Fermi surface and can therefore contribute to low frequency intraband optical responses~\cite{chiral_optical, GME1, GME2}. Previous work on biased metallic systems has focused primarily on electro-optic (EO) responses derived from the Berry curvature dipole, which can produce nonlinear optical effects, non-Hermitian corrections to the optical conductivity, and optical gain~\cite{Fu, Rappoport2023,Morgado2024, BCD_Weyl, PhysRevLett.105.026805}. In parallel, Fermi surface magnetic moment textures can give rise to the gyrotropic magnetic effect, a bias independent optical response closely related to natural optical activity~\cite{GME1,GME2, Ma}. More recently, a distinct magnetoelectric EO response was predicted to arise from the simultaneous presence of Berry curvature and magnetic moment, through a tensor $\mathbf{G}$ formed from their product on the Fermi surface~\cite{magnetoelectric_paper, tdbg}.

A central challenge in identifying this magnetoelectric EO response is that several optical and EO effects may coexist in the same material, even within the low frequency intraband regime considered here~\cite{magnetoelectric_paper}. For the metallic responses considered in this work, where the low frequency optical conductivity is dominated by intraband processes, the relevant tensors are the Berry curvature dipole $\mathbf{D}$, the gyrotropic magnetic tensor $\mathbf{K}$, and the magnetoelectric EO tensor $\mathbf{G}$~\cite{magnetoelectric_paper}. The tensors $\mathbf{D}$ and $\mathbf{K}$ are symmetry equivalent rank 2 pseudotensors, while $\mathbf{G}$ is a rank 2 polar tensor~\cite{GME1, GME2}. As a result, crystal symmetry can constrain these tensors differently in the presence of improper symmetries, such as a mirror symmetry. This difference in transformation behavior provides a symmetry-based route to isolating the magnetoelectric EO response: in suitable noncentrosymmetric space groups (SGs), improper symmetries can forbid the $\mathbf{D}$- and $\mathbf{K}$-driven responses, while preserving the $\mathbf{G}$-driven response. Such materials would provide a cleaner experimental route to the magnetoelectric EO effects, because the competing Berry curvature dipole and gyrotropic magnetic contributions would be absent by symmetry.

In this work, we determine the symmetry-constrained tensor forms of $\mathbf{D}$, $\mathbf{K}$, and $\mathbf{G}$ for all noncentrosymmetric three-dimensional SGs using the TENSOR program of the Bilbao Crystallographic Server (BCS)~\cite{TENSOR}. Our exhaustive symmetry classification shows that SGs 174, 187–190, and 215–220 are the only noncentrosymmetric three-dimensional SGs in which the response pseudotensors $\mathbf{D}$ and $\mathbf{K}$ are forbidden, while the polar response tensor $\mathbf{G}$ remains allowed by symmetry. More generally, our tensor shape tables indicate that no noncentrosymmetric SG forces $\mathbf{G}$ to vanish by symmetry, whereas $\mathbf{D}$ and $\mathbf{K}$ can be symmetry-forbidden. The vanishing of the pseudotensors follows from two classes of point group constraints. In SGs 174 and 187--190, a horizontal mirror plane together with a threefold axis eliminates all components of $\mathbf{D}$ and $\mathbf{K}$. In SGs 215--220, the combined constraints of mutually orthogonal twofold axes, a threefold axis, and a mirror plane remove the pseudotensor components while allowing a diagonal $\mathbf{G}$ form.

We then use first-principles calculations combined with Wannier interpolation to test these symmetry predictions and to identify representative material platforms. The materials Te (SG 152), TaAs (SG 109), and BaTe$_3$ (SG 113)~\cite{Ascencio_EO_Tensor_Shapes} illustrate how different allowed forms of $\mathbf{D}$ and $\mathbf{K}$ can occur while preserving a common form of $\mathbf{G}$. The materials GaAs (SG 216), HgTe (SG 216), and TaN (SG 187)~\cite{Ascencio_EO_Tensor_Shapes} provide examples in which $\mathbf{D}$ and $\mathbf{K}$ are symmetry-forbidden, while $\mathbf{G}$ remains allowed and can be finite. Finally, we propose a simple optical experimental geometry in which the distinct EO responses can be selected by the incident light polarization. In the Te example, $s$-polarized light isolates the $\mathbf{G}$-driven magnetoelectric EO current, while $p$-polarized light accesses the $\mathbf{D}$-driven Berry curvature dipole EO current. For circularly polarized light in the same geometry, the $\mathbf{D}$-driven absorption or gain correction is helicity-even, whereas the $\mathbf{G}$-driven contribution produces a bias-induced circular dichroism (CD) with a characteristic $\sin\theta\cos\theta$ angular dependence. This provides a practical route for optically separating these EO responses in noncentrosymmetric metals.

The manuscript is organized as follows. In Sec.~II we summarize the metallic optical and EO response formalism and define the tensors $\mathbf{D}$, $\mathbf{K}$, and $\mathbf{G}$. In Sec.~III we analyze the symmetry constraints on these tensors and identify the SGs where the magnetoelectric EO response associated with $\mathbf{G}$ is symmetry-isolated. In Sec.~IV we present representative first-principles material calculations. In Sec.~V we discuss experimental geometries for probing and separating the $\mathbf{D}$- and $\mathbf{G}$-driven EO effects.
\section{Metallic Optical and Electro-Optic Response Tensors and Associated Effects}

Here, we summarize the response formalism needed to describe metallic optical responses with and without an applied static electric field in time-reversal symmetric systems lacking spatial inversion symmetry. Details of the derivation can be found in our prior work~\cite{magnetoelectric_paper}. We focus on the low-frequency intraband regime, where
\[
\gamma=\frac{1}{\tau}\ll \omega \ll \frac{\epsilon_{\mathrm{gap}}}{\hbar},
\]
with $\tau$ the relaxation time and $\epsilon_{\mathrm{gap}}$ the optical gap for interband transitions. We first consider the unbiased limit, $\mathbf{E}_0=0$, where the current response is determined only by the optical electric and magnetic fields. Within the relaxation-time approximation, the current density induced by monochromatic optical fields is written as
\begin{equation}
J^{\alpha}(\omega)
=
\sigma_E^{\alpha\beta}(\omega) E_\omega^\beta
+
\sigma_B^{\alpha\beta}(\omega) B_\omega^\beta
\end{equation}

The first term describes the electric field driven optical response, while the second term describes the magnetic field driven optical response. In the unbiased limit, the corresponding optical conductivity tensors can be written as
\begin{equation}
\sigma_E^{\alpha\beta}(\omega)
=
\frac{e^2\tau}{1-i\omega\tau}
V^{\alpha\beta}
\label{eq:unbiased_sigma_E_component}
\end{equation}

\begin{equation}
\sigma_B^{\alpha\beta}(\omega)
=
\frac{e}{\hbar}
\frac{i\omega\tau}{i\omega\tau-1}
K^{\alpha\beta}
\label{eq:unbiased_sigma_B_component}
\end{equation}
Here, $e$ denotes the elementary charge and $\omega$ is the frequency of the optical field.

$\sigma_E(\omega)$ is the usual AC Drude conductivity expressed in terms of the Drude weight tensor
$$
V^{\alpha\beta}
=
\sum_{n,\mathbf{k}}
\left(
-\frac{\partial f^0_{n\mathbf{k}}}{\partial \epsilon_{n\mathbf{k}}}
\right)
v^\alpha_{n\mathbf{k}}v^\beta_{n\mathbf{k}}
$$
where $f^0_{n\mathbf{k}}$ is the equilibrium Fermi function for band $n$ at crystal momentum $\mathbf{k}$, and $v^\alpha_{n\mathbf{k}}$ is the $\alpha$-th component of the band velocity. Meanwhile, $\sigma_B(\omega)$ is expressed in terms of the gyrotropic magnetic tensor
\begin{equation}
K^{\alpha\beta}
=
\hbar
\sum_{n,\mathbf{k}}
\left(
-\frac{\partial f^0_{n\mathbf{k}}}{\partial \epsilon_{n\mathbf{k}}}
\right)
v^\beta_{n\mathbf{k}}m^\alpha_{n\mathbf{k}}
\end{equation}
where $m^\alpha_{n\mathbf{k}}$ is the $\alpha$-th component of the magnetic moment of Bloch electrons~\cite{PhysRevLett.95.137205, PhysRevB.74.024408, PhysRevB.59.14915}. $\mathbf{K}$ captures the magnetic moment texture of Bloch electrons on the Fermi surface. The clean intraband optical regime considered here satisfies $\omega\tau\gg 1$, although Eqs.~(2) and (3) retain the full dependence on $\omega\tau$ within the intraband theory. For completeness, we first summarize their formal collision dominated limit, $\omega\tau\ll 1$, before giving the leading collisionless forms relevant to this work.

The different frequency dependences of $\sigma_E(\omega)$ and $\sigma_B(\omega)$ reflect their distinct physical origins. The electric field directly accelerates carriers and therefore produces the usual Drude response, which remains finite in the dc limit. By contrast, the magnetic field driven gyrotropic response arises from the time-dependent modulation of Bloch state energies by the magnetic moment and therefore vanishes for a static magnetic field. In the formal collision dominated limit, $\omega\tau\ll 1$,

$$
\sigma_E^{\alpha\beta}(\omega)\simeq e^2\tau V^{\alpha\beta},
\qquad
\sigma_B^{\alpha\beta}(\omega)\simeq-i\omega\tau\frac{e}{\hbar}K^{\alpha\beta}
$$

In the collisionless regime relevant to this work, $\omega\tau\gg 1$, while remaining below the interband threshold, $\omega\ll\epsilon_{\mathrm{gap}}/\hbar$,

$$
\sigma_E^{\alpha\beta}(\omega)\simeq\frac{ie^2}{\omega}V^{\alpha\beta},
\qquad
\sigma_B^{\alpha\beta}(\omega)\simeq\frac{e}{\hbar}K^{\alpha\beta}
$$

Thus, the electric field response evolves from a dc Drude conductivity to a reactive $1/\omega$ response, whereas the magnetic field response is intrinsically dynamical and approaches the gyrotropic magnetic coefficient in the collisionless regime.

When a static electric field is applied, the magnetic field driven conductivity acquires an additional Fermi surface contribution. This response can be written compactly by introducing the magnetoelectric EO tensor $\mathbf{G}$, which couples the momentum-space Berry curvature and the magnetic moment of Bloch electrons. The resulting magnetic contribution to the optical conductivity is
\begin{equation}
\sigma_B^{\alpha\beta}(\omega)
=
\frac{e}{\hbar}
\frac{i\omega\tau}{i\omega\tau-1}
\left(
K^{\alpha\beta}
-
e\epsilon^{\alpha\gamma\delta}E_0^\delta G^{\gamma\beta}
\right)
\end{equation}
with 
\begin{equation}
G^{\alpha \beta}
=
\sum_{n,\mathbf{k}}
\left(
-\frac{\partial f^0_{n\mathbf{k}}}{\partial\epsilon_{n\mathbf{k}}}
\right)
m^\beta_{n\mathbf{k}}\Omega^\alpha_{n\mathbf{k}}
\end{equation}
where $\Omega^\alpha_{n\mathbf{k}}$ is the $\alpha$-th component of the Berry curvature of Bloch electrons, $\epsilon$ denotes the Levi-Civita symbol, and repeated indices are summed over. The first term in the magnetic contribution to the optical conductivity is bias independent, while the second term is linear in the applied electric field and therefore describes a magnetoelectric EO correction.

The electric field driven conductivity also acquires bias-induced corrections. These corrections can be expressed in terms of the Berry curvature dipole $\mathbf{D}$, together with 
\begin{equation}
\sigma_E^{\alpha\beta}(\omega)
=
\frac{e^2\tau}{1-i\omega\tau}
\left(
V^{\alpha\beta}
+
\frac{e}{\hbar^2}
\epsilon_{\alpha\gamma\delta}
E_0^\delta
D^{\gamma\beta}
\right)
+
\frac{e^3\tau}{\hbar^2}
\epsilon_{\alpha\gamma\beta}
D^{\gamma\delta}
E_0^\delta
\end{equation}

with 
\begin{equation}
D^{\alpha \beta}
=
\sum_{n,\mathbf{k}} f^0_{n\mathbf{k}}
\frac{\partial\Omega^\alpha_{n\mathbf{k}}}{\partial k_{\beta}}
=
\hbar
\sum_{n,\mathbf{k}}
\left(
-\frac{\partial f^0_{n\mathbf{k}}}{\partial\epsilon_{n\mathbf{k}}}
\right)
v^\beta_{n\mathbf{k}}\Omega^\alpha_{n\mathbf{k}}
\end{equation}

The response tensors introduced above can give rise to experimentally measurable optical and EO effects, with representative examples shown schematically in FIG. 1. As illustrated there, $\mathbf{D}$ can give rise to a bias-induced EO effect, $\mathbf{K}$ can contribute to a bias-independent optical effect, and $\mathbf{G}$ can produce a distinct bias-induced EO effect. The $\mathbf{D}$ tensor describes an EO response in which a static electric field produces a non-Hermitian correction to the optical conductivity, enabling polarization dependent optical gain~\cite{Morgado2024}. In FIG. 1a, this is shown for a selected circular polarization. An optical effect associated with $\mathbf{K}$ is the gyrotropic magnetic effect shown schematically in FIG 1b. In the absence of a static electric bias, the magnetic field component of an incident linearly polarized electromagnetic wave induces a current governed by the Fermi surface magnetic moment texture of Bloch electrons. This low frequency form of natural optical activity can rotate the plane of polarization of the transmitted wave relative to the incident wave~\cite{chiral_optical, GME1, GME2}. The $\mathbf{G}$ tensor produces a magnetoelectric EO response in which the applied static electric field couples the Berry curvature and magnetic moment textures~\cite{magnetoelectric_paper}. This electric bias enabled EO response can appear as circular dichroism through a helicity-dependent absorption or gain correction, where left and right circularly polarized light experience different attenuation or amplification (FIG.~1c)~\cite{tdbg,optical_gain}. Having introduced representative optical and EO effects, we now turn to the symmetry constraints that determine which tensor components, and therefore which effects, are allowed in a given crystal.

\begin{figure}[H]
     \centering
     \includegraphics[width=0.9\linewidth]{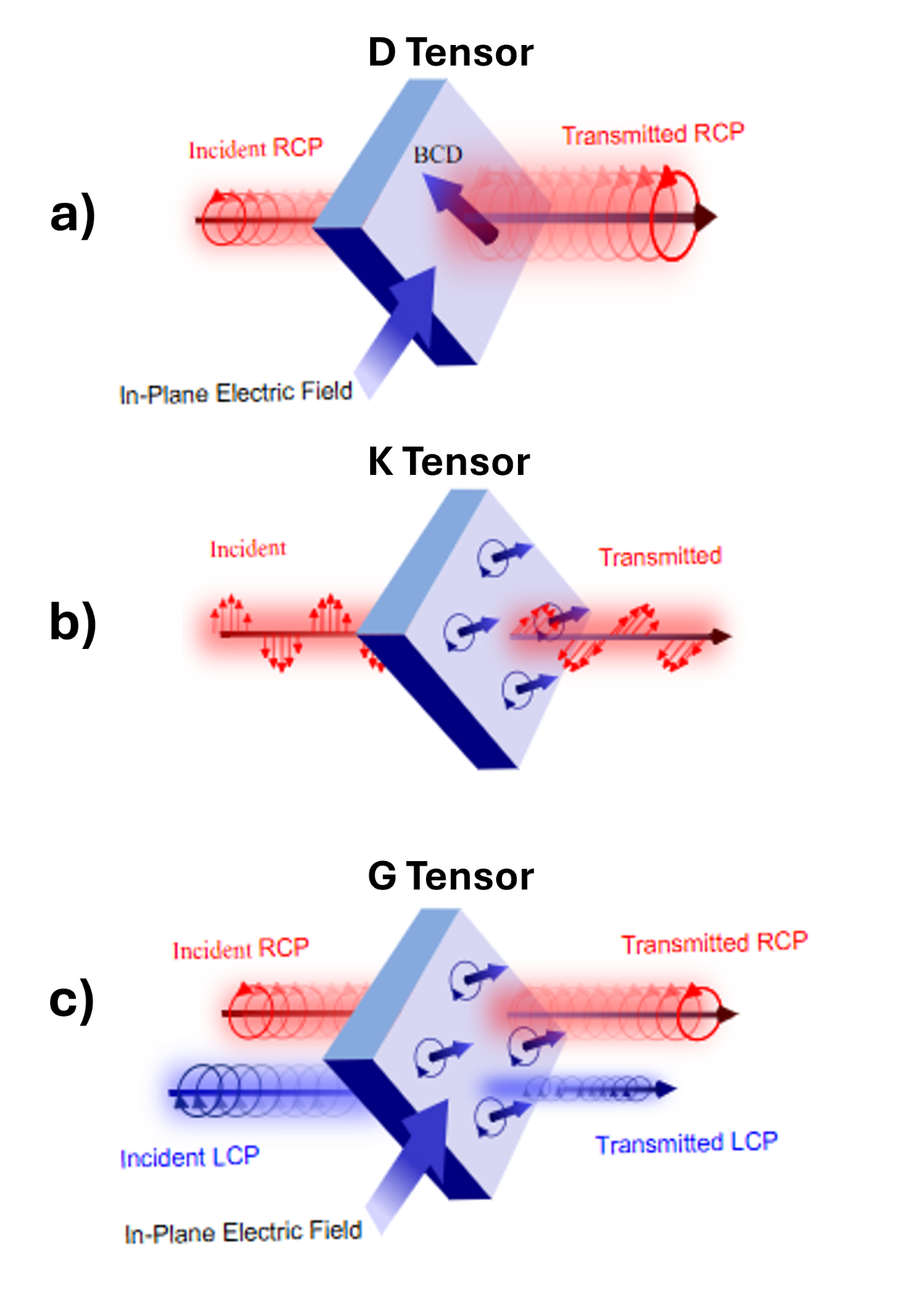}
     \caption{Representative optical and EO effects associated with $\mathbf{D}$, $\mathbf{K}$, and $\mathbf{G}$. a) The Berry curvature dipole $\mathbf{D}$ produces a bias-induced EO response that can generate polarization dependent optical gain. b) The $\mathbf{K}$ tensor is associated with the bias-independent gyrotropic magnetic effect, where a linearly polarized transmitted wave can have its plane of polarization rotated relative to the incident wave. c) The magnetoelectric EO tensor $\mathbf{G}$ produces a bias-induced circular dichroic response, where left and right circularly polarized light are absorbed differently.}
\end{figure}


\section{Effect Of Crystal symmetry on Response Tensor shapes}

Symmetry constraints on $\mathbf{v}_{n\mathbf{k}}$, $\boldsymbol{\Omega}_{n\mathbf{k}}$, and $\mathbf{m}_{n\mathbf{k}}$ directly determine the structure of $\mathbf{D}$, $\mathbf{K}$, and $\mathbf{G}$ (see Eqs. (4), (6), and (8)), and dictate whether the associated optical and EO responses are symmetry-allowed. In the presence of time-reversal symmetry, the $\mathbf{k}$-space Berry curvature and electron magnetic moment, both of which are axial quantities, satisfy
$
\mathbf{A}_{n(-\mathbf{k})} = -\mathbf{A}_{n\mathbf{k}},
$
where $\mathbf{A}_{n\mathbf{k}}$ denotes either $\boldsymbol{\Omega}_{n\mathbf{k}}$ or $\mathbf{m}_{n\mathbf{k}}$. In contrast, spatial inversion symmetry requires
$
\mathbf{A}_{n(-\mathbf{k})} = \mathbf{A}_{n\mathbf{k}}.
$
Thus, in nonmagnetic centrosymmetric materials,
$
\mathbf{A}_{n\mathbf{k}} = -\mathbf{A}_{n\mathbf{k}} = \mathbf{0}
$
at every $\mathbf{k}$ in the Brillouin zone, causing $\mathbf{D}$, $\mathbf{K}$, and $\mathbf{G}$ to vanish identically. Moreover, spatial inversion symmetry alone is sufficient to ensure that both $\mathbf{D}$ and $\mathbf{K}$ vanish due to their symmetry equivalence~\cite{tdbg}. Consequently, optical and EO phenomena arising from $\mathbf{D}$, $\mathbf{K}$, and $\mathbf{G}$ require that the combined constraints of time-reversal and spatial inversion symmetry not both be present~\cite{magnetoelectric_paper}. We therefore focus on time-reversal symmetric systems lacking spatial inversion symmetry, where the metallic optical conductivity is governed by the AC Drude response and the tensors $\mathbf{D}$, $\mathbf{K}$, and $\mathbf{G}$. Determining the symmetry-constrained forms of these tensors then identifies which optical and EO effects are allowed and provides direct guidance for experiment.

Since a periodic potential inherits the symmetries of the crystal, the real-space Hamiltonian remains invariant under the symmetry operations of the crystal SG. A general SG operation is denoted by $\{O_s \mid \mathbf{t}_s\}$, where $O_s$ represents a point group operation and $\mathbf{t}_s$ is a translation vector. The translation part $\mathbf{t}_s$ does not change the transformation of the crystal momentum, which is determined by $O_s$, although it may contribute a phase to the Bloch states~\cite{Martin_2004}. Nevertheless, the full SG determines the little groups and their allowed irreducible representations throughout the Brillouin zone, thereby constraining band hybridization and enforcing certain degeneracies and band connectivity, which can influence the energy dependence and magnitudes of the response tensors. However, for the symmetry constraints on vector functions of $\mathbf{k}$ considered here, only the point group part of the SG operation is relevant to the response tensor shape and the relations among its components. From this point onward, ``symmetries'' refer exclusively to point group symmetries.

The polar nature of $\mathbf{v}_{n\mathbf{k}}$ and the axial nature of $\boldsymbol{\Omega}_{n\mathbf{k}}$ and $\mathbf{m}_{n\mathbf{k}}$ dictate that they transform differently with respect to improper symmetries of the crystal. A point group symmetry represented by an orthogonal matrix $O_s$ imposes the constraint
\begin{equation}
\mathbf{V}_{n\mathbf{k}}
=
\det(O_s) O_s \mathbf{V}_{n(O_s^T\mathbf{k})}
\end{equation}
on an axial vector function $\mathbf{V}_{n\mathbf{k}}$ of the crystal momentum. For a polar vector, the determinant prefactor is omitted. 

The presence of a symmetry represented by $O_s$ guarantees that the eigenenergies of the Bloch Hamiltonian satisfy
\begin{equation}
\varepsilon_{n\mathbf{k}}
=
\varepsilon_{n(O_s^T\mathbf{k})},
\end{equation}
provided the band label is chosen consistently under the symmetry. It follows that
\begin{equation}
\frac{\partial f^0_{n\mathbf{k}}}{\partial \varepsilon_{n\mathbf{k}}}
=
\frac{\partial f^0_{n\mathbf{k}'}}{\partial \varepsilon_{n\mathbf{k}'}},
\qquad
\mathbf{k}'=O_s^T\mathbf{k}.
\end{equation}
Moreover, the Brillouin zone maps onto itself under point group symmetry operations, so the substitution $\sum_{\mathbf{k}}\rightarrow \sum_{\mathbf{k}'}$ can be made.

Material systems of interest are those that possess time-reversal symmetry, so we focus on noncentrosymmetric SGs to ensure that the $\mathbf{k}$-space Berry curvature and magnetic moment of Bloch electrons are allowed by symmetry. To determine the shapes of the $\mathbf{D}$, $\mathbf{K}$, and $\mathbf{G}$ tensors in three-dimensional crystals, we employ the TENSOR program available through the Bilbao Crystallographic Server (BCS)~\cite{TENSOR}. We adopt the BCS convention in which each space group is represented in its standard crystallographic setting, while tensor components are expressed in the associated right handed orthogonal reference frame. This frame is defined by $\mathbf{a}'\parallel\mathbf{a}$, $\mathbf{c}'\parallel\mathbf{c}^{*}$, and $\mathbf{b}'\parallel\mathbf{c}'\times\mathbf{a}'$, where the unprimed vectors refer to the conventional crystallographic basis and the primed vectors define the associated orthogonal BCS reference frame. Here, $\mathbf{a}$ is a conventional direct lattice vector and $\mathbf{c}^{*}$ is the corresponding reciprocal lattice vector. Accordingly, the tensor indices $x$, $y$, and $z$ refer to the orthogonal directions $\mathbf{a}'$, $\mathbf{b}'$, and $\mathbf{c}'$, respectively.

The tensor tables in the Appendix are intended to serve as a symmetry guide for material and experimental geometry selection. Given a candidate material, one first identifies its SG and then reads off the allowed components of the pseudotensors $\mathbf{D}$ and $\mathbf{K}$ and the polar tensor $\mathbf{G}$. The table immediately shows whether a given optical or EO mechanism is symmetry-forbidden, whether different tensor components are independent or symmetry-related, and whether a material belongs to the symmetry-isolated class in which $\mathbf{D}$ = $\mathbf{K} = \mathbf{0}$ while $\mathbf{G} \neq \mathbf{0}$ remains allowed. The point-group-property column provides additional context: the proper/improper label indicates whether the polar-versus-pseudotensor distinction can affect the tensor shape, while the polar/nonpolar label indicates whether the crystal admits a symmetry-allowed polar axis that may be useful for choosing bias-field or measurement directions. The tables determine only the symmetry-allowed tensor structure; the magnitudes and signs of the surviving components remain material-specific Fermi-surface quantities and must be obtained from first-principles calculations or experiment. 

To begin, we perform a symmetry analysis on \textbf{D}, \textbf{K}, and \textbf{G} tensors for cases where the shape of a specific tensor type is identical across a large number of SGs in order to elucidate some of the underlying symmetries that enforce such shapes. 

\subsection{Symmetry constraints on $\mathbf{D}$, $\mathbf{K}$, and $\mathbf{G}$}
As noted above, $\mathbf{D}$ and $\mathbf{K}$ are symmetry equivalent. Therefore, they transform identically under each symmetry operation and are subject to the same tensor constraints. The tensors $\mathbf{D}$ and $\mathbf{K}$ are rank-2 pseudotensors, and their intrinsic tensor symmetry is encoded by the Jahn symbol $eV^2$~\cite{TENSOR, Lannebere2025Symmetry}. By contrast, $\mathbf{G}$ is defined from a product of two axial vector components, so it transforms as a rank-2 polar tensor and is described by the Jahn symbol $V^2$. These Jahn symbols are used as input to the TENSOR program to determine the corresponding symmetry-constrained tensor forms for all 138 noncentrosymmetric SGs.

Among the noncentrosymmetric SGs, $\mathbf{D}$ and $\mathbf{K}$ are constrained to diagonal forms in 51 cases (SGs 16--24, 89--98, 111--114, 121--122, 149--155, 177--182, 195--199, and 207--214), while $\mathbf{G}$ is diagonal in a larger set of 111 cases (SGs 16--46, 89--122, 149--161, 177--190, 195--199, and 207--220). For $\mathbf{D}$ and $\mathbf{K}$, these diagonal forms are associated with either two mutually orthogonal twofold axes or a threefold axis together with an orthogonal twofold axis (see Supplementary Materials~\cite{SM}).  For $\mathbf{G}$, additional point groups containing mirror symmetries also allow diagonal tensor forms~\cite{SM}. Other recurring shapes for $\mathbf{D}$ and $\mathbf{K}$ involve the vanishing of the diagonal components of these optical and EO response tensors (SGs 6--9, 25--46, 99--110, 115--120, 156--161, and 183--186). In such cases it is an appropriately oriented mirror plane symmetry that enforces these components to vanish~\cite{SM}. Additional symmetries may further force other tensor components to vanish or interrelate otherwise independent ones.

In the SGs discussed so far, all three tensors are symmetry-allowed. To experimentally isolate the recently predicted magnetoelectric EO effects associated with $\mathbf{G}$, however, it is useful to identify cases in which $\mathbf{D}$ and $\mathbf{K}$ vanish by symmetry. Then the leading order EO response is governed by the \emph{coupling} between the Berry curvature and the magnetic moment of Bloch electrons encoded in $\mathbf{G}$. We identify 11 noncentrosymmetric SGs that satisfy this condition and discuss the symmetries responsible for the vanishing of $\mathbf{D}$ and $\mathbf{K}$.

\subsection{Symmetry-isolation of the $\mathbf{G}$-Driven Electro-Optic Effects}

We now focus on this symmetry-isolated set, where the magnetoelectric EO response associated with $\mathbf{G}$ remains allowed, while the EO contribution from $\mathbf{D}$ and the optical response associated with $\mathbf{K}$ are forbidden.

Of the 11 SGs of interest (SGs 174, 187--190, and 215--220), five belong to the hexagonal crystal family. In SGs 174 and 187--190, the absence of $\mathbf{D}$ and $\mathbf{K}$ follows from the presence of a threefold axis and an orthogonal mirror plane. The remaining six SGs, SGs 215--220, belong to the cubic crystal family. In these cases, the same tensors are forbidden by the combined constraints of a threefold axis along a body diagonal, two mutually orthogonal twofold axes, and an appropriately oriented mirror plane symmetry.

We next explain how the relevant symmetry constraints eliminate $\mathbf{D}$ and $\mathbf{K}$ in these two classes of SGs. For SGs 174 and 187--190, the relevant generators may be chosen as a horizontal mirror plane, $M_z$, and a threefold axis, $C_{3z}$. Under
$M_z:(x,y,z)\rightarrow(x,y,-z)$,
$$
\begin{aligned}[t]
v^x &\rightarrow v^x, \qquad
v^y \rightarrow v^y, \qquad
v^z \rightarrow -v^z, \\[-0.2em]
\Omega^x &\rightarrow -\Omega^x, \qquad
\Omega^y \rightarrow -\Omega^y, \qquad
\Omega^z \rightarrow \Omega^z .
\end{aligned}
$$
Substituting these transformation rules into the definition of $\mathbf{D}$ given in Section II, using
$\varepsilon_{n\mathbf{k}}=\varepsilon_{n(M_z^{-1}\mathbf{k})}$, and changing variables in the Brillouin zone sum gives
$$
\mathbf{D}
=
\begin{pmatrix}
0 & 0 & D^{xz} \\
0 & 0 & D^{yz} \\
D^{zx} & D^{zy} & 0
\end{pmatrix}.
$$
Thus, $M_z$ removes all diagonal components of $\mathbf{D}$, leaving only $D^{xz}$, $D^{yz}$, $D^{zx}$, and $D^{zy}$. These remaining components are then forbidden by the $C_{3z}$ constraint,
$$
D^{xz}=D^{yz}=D^{zx}=D^{zy}=0.
$$
This follows from the same $C_{3z}$ argument used for $\mathbf{G}$ in the Supplementary Materials~\cite{SM}. Since $C_{3z}$ is a proper operation, polar and axial vectors transform in the same way under this rotation, so the corresponding argument also applies to $\mathbf{D}$. Therefore, $\mathbf{D}$ vanishes identically in SGs 174 and 187--190. Since $\mathbf{K}$ is symmetry equivalent to $\mathbf{D}$, the same conclusion applies to $\mathbf{K}$. The tensor $\mathbf{G}$ is not forbidden by these same constraints, because it is a rank-2 polar tensor formed from the product of two axial quantities. Under $M_z$, the diagonal products $m^\alpha\Omega^\alpha$ are even, so this mirror symmetry does not force the diagonal components of $\mathbf{G}$ to vanish.

For SGs 215--220, a convenient set of generators is given by $C_{2z}$, $C_{2y}$, $C_{3[111]}$, and $M_{[1\bar{1}0]}$. As shown in the Supplementary Materials~\cite{SM}, the presence of two mutually orthogonal twofold axes forces all off-diagonal components of $\mathbf{D}$ to vanish. In addition, as shown in our prior work~\cite{SG198_EO}, the $C_{3[111]}$ symmetry relates the diagonal components according to $D^{xx}=D^{yy}=D^{zz}$. The final constraint comes from the mirror plane $M_{[1\bar{1}0]}$. Both $v^z$ and $\Omega^z$ lie parallel to this mirror plane. However, $v^z$ is polar while $\Omega^z$ is axial. Their product is therefore odd under the mirror symmetry, forcing $D^{zz}=0$. Together with the $C_{3[111]}$ constraint, this gives $D^{xx}=D^{yy}=D^{zz}=0$. Hence, all components of $\mathbf{D}$ vanish in SGs 215--220, and the same result follows for $\mathbf{K}$. In contrast, the diagonal products $m^\alpha \Omega^\alpha$ entering $\mathbf{G}$ are even under $M_{[1\bar{1}0]}$, so this mirror symmetry does not forbid the diagonal components of $\mathbf{G}$.

Having established the symmetry conditions under which the response tensors are allowed or forbidden, we now test these predictions using first-principles calculations for representative material systems. In the following section, we present density functional theory (DFT)-based results for six materials spanning different noncentrosymmetric SGs. These examples demonstrate that the numerically calculated tensor components are consistent with the symmetry-constrained forms identified in this work.

\section{Material Examples}

Because $\mathbf{D}$ and $\mathbf{K}$ are pseudotensors, their symmetry constrained forms exhibit more variation than that of the polar tensor $\mathbf{G}$. This provides additional flexibility in designing experiments. In particular, one can select materials from SGs that share the same allowed form of $\mathbf{G}$, while realizing different allowed forms of $\mathbf{D}$ and $\mathbf{K}$. As a result, a fixed experimental geometry may couple to the same components of $\mathbf{G}$ across multiple materials, while allowing the contributions from different components of $\mathbf{D}$ and $\mathbf{K}$ to be compared or isolated through materials selection.

In FIG.~2, we show the energy dependence of the three response tensors for Te (SG 152), TaAs (SG 109), and BaTe$_3$ (SG 113), obtained from DFT-based calculations combined with Wannier interpolation~\cite{QE,W90, TaAs, GME2, AHE_Wannier}. The resulting tensor components are consistent with the symmetry constrained forms listed in the Appendix and demonstrate how different material choices can realize distinct combinations of the allowed responses. Importantly, these examples all allow the same diagonal form of $\mathbf{G}$, while the allowed forms of $\mathbf{D}$ and $\mathbf{K}$ differ from material to material.

For Te, all three tensors are diagonal with $c_{xx}=c_{yy}$ and an independent $c_{zz}$ component, as expected for SG 152. This is displayed in FIG.~2(a), where the off-diagonal components vanish and the $xx$ and $yy$ components overlap. In contrast, for TaAs (SG 109), symmetry allows only the $xy$ and $yx$ components of $\mathbf{D}$ and $\mathbf{K}$, with $D^{xy}=-D^{yx}$ and $K^{xy}=-K^{yx}$. This is reflected in FIG.~2(b), while $\mathbf{G}$ remains diagonal. For BaTe$_3$, FIG.~2(c) shows the SG 113 form, where $\mathbf{D}$ and $\mathbf{K}$ have diagonal components with $c_{xx}=-c_{yy}$, while $\mathbf{G}$ again has the same diagonal form.

Beyond confirming the allowed tensor components, FIG.~2 also shows that the response magnitudes can vary strongly with energy. This is important because all three tensors are Fermi surface quantities, so their values can be tuned by shifting the Fermi level. For gapped materials such as Te and BaTe$_3$, useful optical or EO signals will require carrier doping or gating to move the Fermi level into the valence or conduction band where the allowed tensor components are appreciable. 
\begin{figure}[H]
     \centering
     \includegraphics[width=0.95\linewidth]{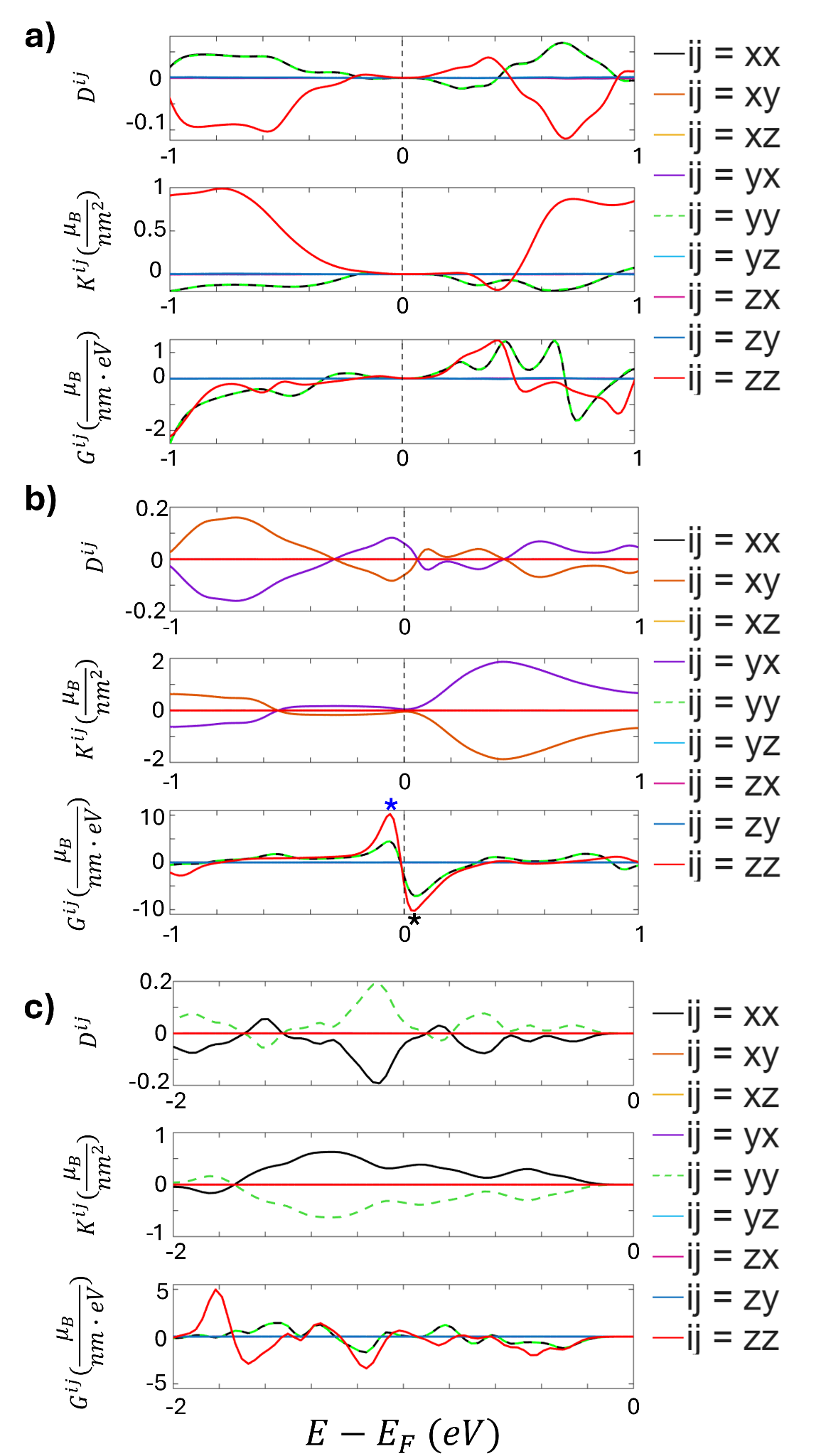}
     \caption{Energy dependence of the $\mathbf{D}$, $\mathbf{K}$, and $\mathbf{G}$ tensors for a) Te (SG 152), b) TaAs (SG 109), and c) BaTe$_3$ (SG 113), illustrating how materials selection can modify the allowed $\mathbf{D}$ and $\mathbf{K}$ responses while preserving a common form of $\mathbf{G}$. In the TaAs $\mathbf{G}$ panel, the blue and black asterisks indicate the nearby peak and dip features discussed in the text, located near $E-E_F=-0.06$ eV and $E-E_F=0.045$ eV, respectively.}
\end{figure}

Among the three examples, TaAs (FIG. 2b) shows the largest overall response magnitudes for $\mathbf{D}$, $\mathbf{K}$, and $\mathbf{G}$. Because TaAs is metallic, all three tensor coefficients are already finite at the equilibrium Fermi level, so no Fermi level shift is required to access these responses. In addition, the $\mathbf{G}$ tensor exhibits a sharp peak and dip very close to $E_F$, at approximately $E-E_F=-0.06$ eV (blue asterisk) and $E-E_F=0.045$ eV (black asterisk). At $E-E_F=-0.06$ eV, the two independent $\mathbf{G}$ components are $G^{xx}=G^{yy}\simeq 4.46\,\mu_B/(\mathrm{nm}\cdot\mathrm{eV})$ and $G^{zz}\simeq 10.25\,\mu_B/(\mathrm{nm}\cdot\mathrm{eV})$, while at $E-E_F=0.045$ eV they are $G^{xx}=G^{yy}\simeq -7.16\,\mu_B/(\mathrm{nm}\cdot\mathrm{eV})$ and $G^{zz}\simeq -10.24\,\mu_B/(\mathrm{nm}\cdot\mathrm{eV})$. Since $E-E_F$ corresponds to a shift of the Fermi level relative to its equilibrium value, these nearby features may be accessible through modest Fermi level tuning. Importantly, the corresponding allowed components of $\mathbf{D}$ and $\mathbf{K}$ are relatively small at these nearby Fermi level shifts: at $E-E_F=-0.06$ eV, $\left|D^{yx(xy)}\right|\simeq 0.083$ and $\left|K^{yx(xy)}\right|\simeq 0.102\,\mu_B/\mathrm{nm}^2$, while at $E-E_F=0.045$ eV, $\left|D^{yx(xy)}\right|\simeq 0.019$ and $\left|K^{yx(xy)}\right|\simeq 0.087\,\mu_B/\mathrm{nm}^2$. Thus, TaAs provides an example where the metallic response is already finite at the undoped Fermi level, while modest Fermi level tuning can further enhance the desired $\mathbf{G}$ contribution relative to the competing $\mathbf{D}$ and $\mathbf{K}$ responses.

These examples illustrate how the tensor forms obtained from symmetry and the energy dependent calculations play complementary roles. The symmetry analysis identifies which tensor components are allowed in a given material, while the first-principles results show how the responses evolve as the Fermi level is shifted. This provides a practical route for choosing materials and experimentally relevant Fermi level positions to enhance or suppress the relative contributions from $\mathbf{D}$, $\mathbf{K}$, and $\mathbf{G}$ in a chosen experimental geometry.

As discussed above, materials in which $\mathbf{D}$ and $\mathbf{K}$ vanish are of experimental interest because they provide a clearer route to isolating EO effects associated with the newly predicted $\mathbf{G}$ tensor. To numerically confirm that $\mathbf{D}$ and $\mathbf{K}$ are symmetry-forbidden in the SGs of interest while $\mathbf{G}$ can be nonzero, we calculate the energy dependence of $\mathbf{G}$ in FIG. 3 and show that $\mathbf{D}$ and $\mathbf{K}$ vanish up to numerical precision~\cite{SM}.

For GaAs and HgTe, which both belong to SG 216, the calculated $\mathbf{G}$ tensor has the cubic form with $G^{xx}=G^{yy}=G^{zz}$, while all off-diagonal components vanish. This is consistent with the isotropic form allowed for $\mathbf{G}$ in SGs 215--220 in the Appendix. In GaAs (FIG. 3a), the diagonal components show appreciable features below the reference Fermi level, particularly near the valence band edge. For example, at $E-E_F=-0.2402$ eV (blue asterisk), the isotropic component reaches $G^{xx}=G^{yy}=G^{zz}\simeq -1.108\,\mu_B/(\mathrm{nm}\cdot\mathrm{eV})$. This suggests that lowering the Fermi level toward the valence band could enhance the $\mathbf{G}$-driven response in GaAs. HgTe shows the same isotropic tensor shape with diagonal components overlapping over the plotted range, again confirming the cubic symmetry constraint. Compared with GaAs, the response varies more smoothly across the displayed energy window, suggesting that moderate Fermi level tuning may be sufficient to access a finite magnetoelectric EO signal.

\begin{widetext}
\begin{center}
\includegraphics[width=0.65\textwidth]{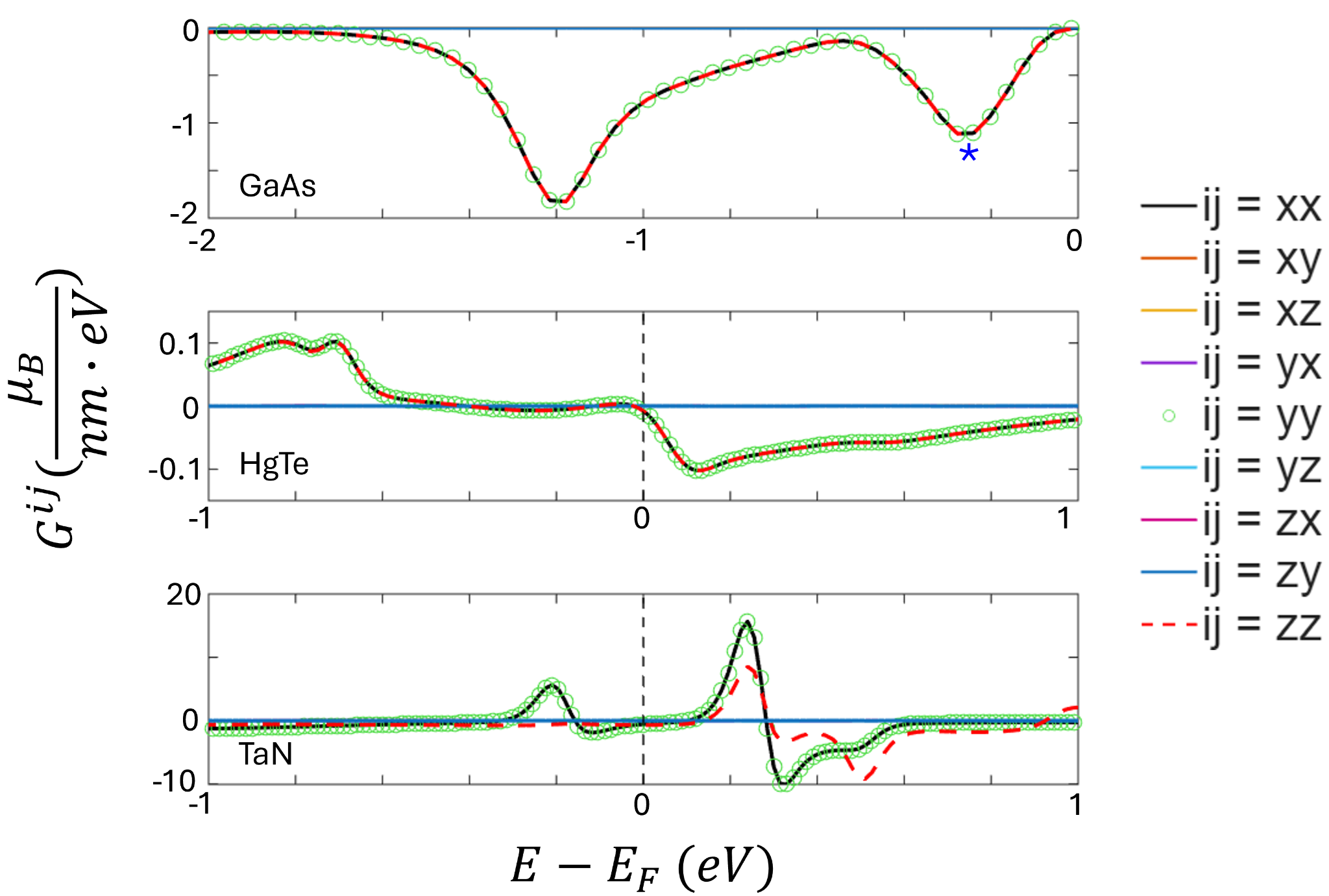}

\vspace{0.5em}
\refstepcounter{figure}
\label{fig:G_tensor_energy}
\begin{minipage}{0.85\textwidth}
\small
\textbf{FIG.~\thefigure.} Energy dependent $\mathbf{G}$ tensor for GaAs (SG 216), HgTe (SG 216), and TaN (SG 187). The blue asterisk in the GaAs panel marks the feature discussed in the text, located near $E-E_F=-0.2402$ eV, where the isotropic component reaches $G^{xx}=G^{yy}=G^{zz}\simeq -1.108\,\mu_B/(\mathrm{nm}\cdot\mathrm{eV})$.
\end{minipage}
\end{center}
\end{widetext}

TaN belongs to SG 187, one of the hexagonal SGs in which $\mathbf{D}$ and $\mathbf{K}$ vanish while $\mathbf{G}$ is allowed. The calculated $\mathbf{G}$ tensor is diagonal with $G^{xx}=G^{yy}$ and an independent $G^{zz}$ component, while the off-diagonal components vanish. The allowed components show large features near the equilibrium Fermi level. In particular, $G^{xx}=G^{yy}$ reaches approximately $5.63\,\mu_B/(\mathrm{nm}\cdot\mathrm{eV})$ at $E-E_F=-0.21$ eV and $15.66\,\mu_B/(\mathrm{nm}\cdot\mathrm{eV})$ at $E-E_F=0.24$ eV. The response also changes sign near these features, so small Fermi level shifts can modify both the magnitude and sign of the magnetoelectric EO response. Thus, TaN provides a useful example in which a large $\mathbf{G}$ response is experimentally accessible while the competing $\mathbf{D}$ and $\mathbf{K}$ responses are forbidden by symmetry.

Taken together, FIG. 3 confirms the key symmetry prediction of the previous section: in the selected SGs, $\mathbf{D}$ and $\mathbf{K}$ are forbidden, while $\mathbf{G}$ remains allowed and can vary strongly with energy. These material examples therefore show that the magnetoelectric EO response can be investigated without being obscured by the Berry curvature dipole and gyrotropic magnetic contributions. We next show how the distinct EO responses can be accessed in a simple optical experimental configuration.
\section{Experimental Design for Disentangling Distinct EO Contributions}
As an example of a possible experimental design, we consider an experiment based on GaAs (SG 216, $\epsilon^{intrin}_{\mathrm{gap}} \simeq 1.42\text{eV}~ \text{or}~ f^{intrin}_{max} \approx 343\text{THz}$). Although other materials discussed above also have nonzero $\mathbf{G}$ tensors and may exhibit larger $\mathbf{G}$-driven responses, GaAs provides a simple and experimentally accessible platform for illustrating how the symmetry-isolated $\mathbf{G}$ response could be probed in a practical experimental setup. For this material, the $\mathbf{G}$ tensor has the isotropic form
\[
\mathbf{G}
=
\begin{pmatrix}
G & 0 & 0 \\
0 & G & 0 \\
0 & 0 & G
\end{pmatrix}
\]
consistent with FIG. 3. We take the crystal surface to lie in the $x$-$z$ plane, so that the surface normal points along $-y$. Linearly polarized light is incident in the $z=0$ plane at an angle $\theta$ measured with respect to the $+y$ axis, and the static electric bias field is applied along $+z$, as shown in FIG. 4a. Assuming these experimental conditions and plugging this $\mathbf{G}$ into the expression for the EO conductivity (see (5)) expressed in the basis of the optical electric field gives

\begin{equation}
\bar{\boldsymbol{\sigma}}^{\mathrm{EO}}(\omega)
=
\frac{e^2}{c\hbar}
E_0^z
\frac{i\omega\tau}{i\omega\tau-1}
G
\begin{pmatrix}
0 & 0 & \sin\theta \\
0 & 0 & \cos\theta \\
0 & 0 & 0
\end{pmatrix}
\end{equation}

We now consider two special cases. For $s$-polarized light, the optical electric field is perpendicular to the plane of incidence,
$$
\mathbf{E}_{\omega}^{s}
=
E_{\omega}^{s}
\begin{pmatrix}
0\\
0\\
1
\end{pmatrix}$$

so the EO current density is
\begin{equation}
\mathbf{J}_{\mathrm{EO}}^{s}(\omega)
=
\bar{\boldsymbol{\sigma}}^{\mathrm{EO}}(\omega)
\mathbf{E}_{\omega}^{s}
=
\frac{e^2}{c\hbar}
E_0^z
\frac{i\omega\tau}{i\omega\tau-1}
G E_{\omega}^{s}
\begin{pmatrix}
\sin\theta\\
\cos\theta\\
0
\end{pmatrix}
\end{equation}

For $p$-polarized light, the optical electric field lies in the plane of incidence and is perpendicular to $\mathbf{k}=(\sin\theta,\cos\theta,0)$,
$$
\mathbf{E}_{\omega}^{p}
=
E_{\omega}^{p}
\begin{pmatrix}
\cos\theta\\
-\sin\theta\\
0
\end{pmatrix}
$$
which gives
\begin{equation}
\mathbf{J}_{\mathrm{EO}}^{p}(\omega)
=
\bar{\boldsymbol{\sigma}}^{\mathrm{EO}}(\omega)
\mathbf{E}_{\omega}^{p}
=
\begin{pmatrix}
0\\
0\\
0
\end{pmatrix}
\end{equation}
Thus, in this geometry the GaAs EO current arises only for $s$-polarized light, since $\mathbf{D}$ is absent by symmetry and the $\mathbf{G}$-driven conductivity couples only to the $z$-component of the optical electric field. We now ask whether a suitable choice of geometry can isolate the $\mathbf{G}$-driven EO response even when $\mathbf{D}$ and $\mathbf{K}$ are symmetry-allowed.

We take Te (SG 152, $\epsilon^{intrin}_{\mathrm{gap}} \simeq 0.33\text{eV}~ \text{or}~ f^{intrin}_{max} \approx 80\text{THz}$) as an example. The EO response tensors are given by
\begin{equation}
\mathbf{D}
=
\begin{pmatrix}
D_{\perp} & 0 & 0 \\
0 & D_{\perp} & 0 \\
0 & 0 & D_z
\end{pmatrix},
\qquad
\mathbf{G}
=
\begin{pmatrix}
G_{\perp} & 0 & 0 \\
0 & G_{\perp} & 0 \\
0 & 0 & G_z
\end{pmatrix}
\end{equation}
We use the same geometry as in the GaAs experiment discussed above (FIG. 4a) and specialize to the case of $s$-polarized and $p$-polarized light. The resulting EO current densities are

\begin{equation}
\mathbf{J}^{\mathrm{EO},s}_{\omega}
=
\frac{e^2}{c\hbar}
E_0^z
\frac{i\omega\tau}{i\omega\tau-1}
G_{\perp}E_{\omega}^{s}
\begin{pmatrix}
\sin\theta\\
\cos\theta\\
0
\end{pmatrix}
\end{equation}

\begin{equation}
\mathbf{J}^{\mathrm{EO},p}_{\omega}
=
\frac{e^3\tau}{\hbar^2}
E_0^z
\left(
D_z
-
\frac{D_{\perp}}{1-i\omega\tau}
\right)
E_{\omega}^{p}
\begin{pmatrix}
\sin\theta\\
\cos\theta\\
0
\end{pmatrix}
\end{equation}

This example shows that symmetry isolation is not the only way to access the $\mathbf{G}$-driven EO effects. Even in materials such as Te, where both $\mathbf{D}$ and $\mathbf{G}$ are symmetry-allowed, the experimental geometry can be chosen so that the $\mathbf{D}$-driven EO current is zero. For the geometry considered here, $s$-polarized light has $\mathbf{E}_{\omega}\parallel z$, while the two $\mathbf{D}$ associated terms act only on the $x$ and $y$ components of the optical electric field. As a result, the $\mathbf{D}$ contribution vanishes for $s$-polarized light, whereas the $\mathbf{G}$ term remains finite. Thus, an appropriate choice of polarization, bias direction, and incidence plane can isolate the magnetoelectric EO response even in materials where $\mathbf{D}$ and $\mathbf{K}$ are not symmetry-forbidden.

Importantly, for the experimental geometry below (see FIG.~4), the $\mathbf{G}$-driven EO current for $s$-polarized light is proportional to $(\sin\theta,\cos\theta,0)^T$, so at normal incidence the response is directed along the surface normal and its tangential component vanishes. Therefore, a CD signal arises from the tangential EO current at oblique incidence. To see this explicitly, we write circularly polarized light in the $p/s$ basis as
\[
\mathbf{E}^{\eta}_{\omega}
=
\frac{E_{\omega}}{\sqrt{2}}
\left(
\mathbf{e}_p+i\eta\mathbf{e}_s
\right),
\qquad
\eta=\pm 1,
\]
so that $E_{\omega}^{p}=E_{\omega}/\sqrt{2}$ and $E_{\omega}^{s}=i\eta E_{\omega}/\sqrt{2}$. Since the surface normal is along $-y$, the relevant tangential EO current is the $x$ component,
\[
J_{x,\omega}^{\mathrm{EO},\eta}
=
\frac{E_{\omega}}{\sqrt{2}}
E_0^z \sin\theta
\left[
f_{D}(D_{z},D_{\perp};\omega)
+
f_{G}(G_{\perp};\omega, \eta)
\right]
\]

where
\[
f_{D}(D_{z},D_{\perp};\omega)
\equiv
\frac{e^{3}\tau}{\hbar^{2}}
\left(
D_{z}
-
\frac{D_{\perp}}{1-i\omega\tau}
\right)
\]
and
\[
f_{G}(G_{\perp};\omega, \eta)
\equiv
i\eta\frac{e^{2}}{c\hbar}
\frac{i\omega\tau}{i\omega\tau-1}
G_{\perp}.
\]

The helicity-dependent EO absorption correction is proportional to
\[
\begin{aligned}
A_{\eta}^{\mathrm{EO}}
&\propto
\frac{|E_{\omega}|^{2}}{2}
E_{0}^{z}
\sin\theta\cos\theta \\
&\quad\times
\operatorname{Re}
\left[
f_{D}(D_{z},D_{\perp};\omega)
+
f_{G}(G_{\perp};\omega,\eta)
\right]
\end{aligned}
\]

The $\mathbf{D}$-driven term is independent of the handedness $\eta$, while the $\mathbf{G}$-driven term changes sign between LCP and RCP. Thus, the circular dichroism isolates the $\mathbf{G}$-driven contribution,

\begin{equation}
\mathrm{CD}
\propto
|E_{\omega}|^2
E_0^z
\sin\theta\cos\theta
\frac{e^2}{c\hbar}
G_{\perp}
\frac{\omega\tau}{1+(\omega\tau)^2}
\end{equation}
This bias-induced CD signal vanishes at $\theta =0^\circ$ and $\theta = 90^\circ$ , and is maximal in magnitude at $\theta=45^\circ$.

\begin{figure}[H]
     \centering
          \includegraphics[width=0.8\linewidth]{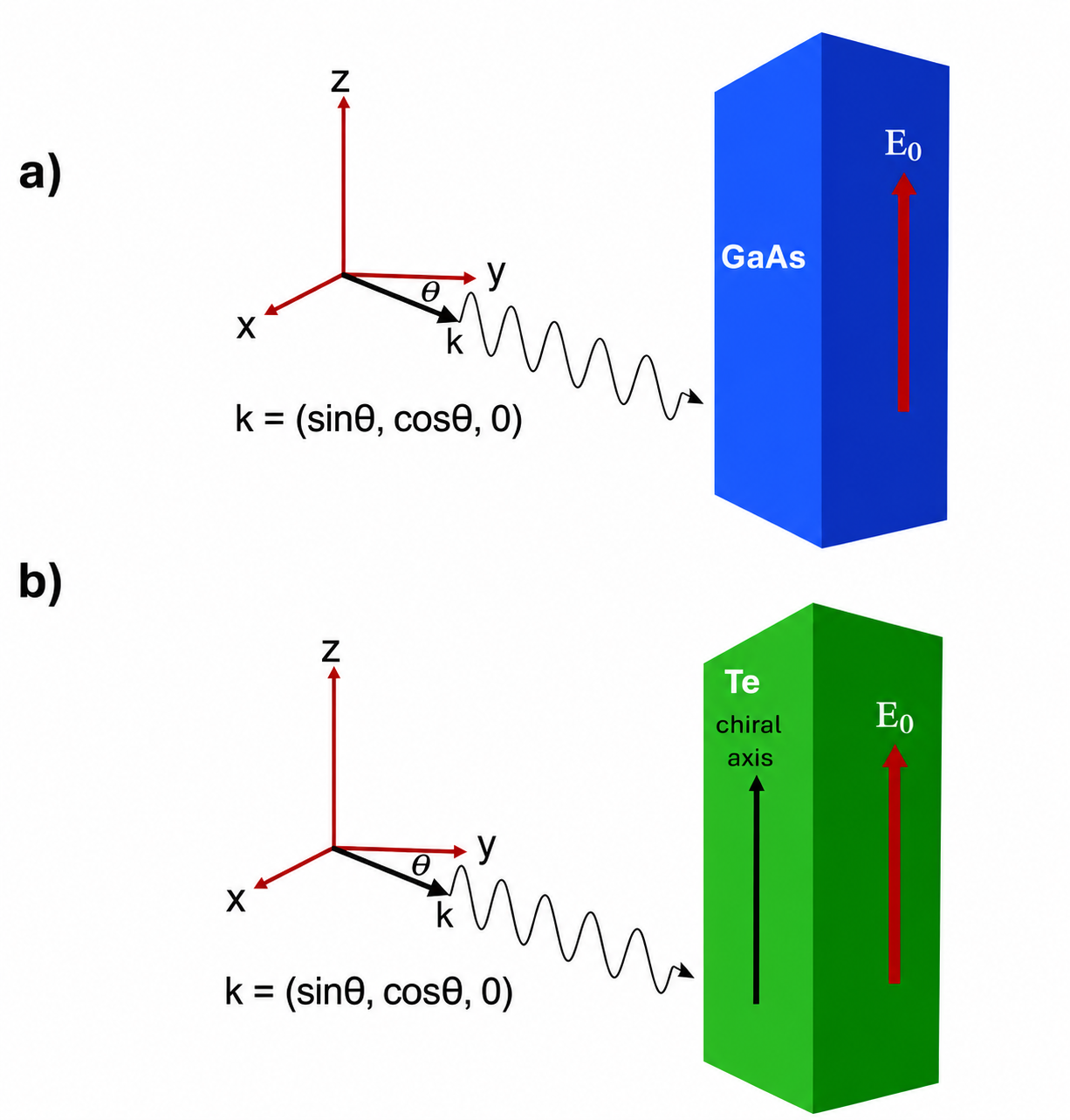}
      \caption{Schematic experimental setup for probing the $\mathbf{G}$-driven EO response. a) GaAs geometry with the crystal surface in the $xz$ plane, surface normal along $-y$, optical wave incident in the $z=0$ plane with wave vector $\mathbf{k}=(\sin\theta,\cos\theta,0)$, and static electric bias field $\mathbf{E}_0$ applied along $+z$. b) Te in the same experimental geometry, with the chiral axis aligned along $+z$.}
\end{figure}

From a fundamental perspective, this experimental configuration can selectively probe the Fermi surface Berry curvature texture encoded in the Berry curvature dipole $\mathbf{D}$ and the combined Berry curvature and magnetic moment textures encoded in the magnetoelectric tensor $\mathbf{G}$ through an appropriate choice of incident polarization. From a practical perspective, these results highlight the broader range of functionalities that noncentrosymmetric metals can support, including polarization selective optical amplification and bias induced circular dichroism, which may provide a route toward next generation optical modulators.

\section{Conclusions}

We classified the symmetry-constrained forms of the Berry curvature dipole $\mathbf{D}$, gyrotropic magnetic tensor $\mathbf{K}$, and magnetoelectric electro-optic (EO) tensor $\mathbf{G}$ for all noncentrosymmetric three dimensional space groups (SGs). This analysis identifies 11 SGs, SGs 174, 187--190, and 215--220, in which the magnetoelectric EO response described by $\mathbf{G}$ should be more experimentally accessible. In these SGs, the competing Berry curvature dipole EO response from $\mathbf{D}$ and gyrotropic magnetic response from $\mathbf{K}$ vanish by symmetry, while $\mathbf{G}$ remains allowed because of their different transformation properties under improper operations.

First principles calculations confirmed the predicted tensor forms and demonstrated that the magnitude and sign of the allowed EO responses can be tuned by shifting the Fermi level. GaAs (SG 216), HgTe (SG 216), and TaN (SG 187) were identified as representative materials in which the $\mathbf{G}$ contribution remains symmetry-allowed and can be finite, while the $\mathbf{D}$ and $\mathbf{K}$ contributions vanish by symmetry. For Te (SG 152), we showed that the distinct EO mechanisms can instead be separated through their polarization and angular dependence: the $\mathbf{D}$ contribution is helicity independent, whereas the $\mathbf{G}$ contribution produces a bias induced circular dichroism proportional to $\sin\theta\cos\theta$, which vanishes at normal and grazing incidence and is maximal at $\theta=45^\circ$. These results highlight how the choice of SG and experimental geometry provides complementary strategies for probing magnetoelectric EO effects in noncentrosymmetric metals.

\textit{Acknowledgments}. The authors thank Tom Folland and Seon Namgung for helpful discussions regarding the experimental design. C.O.A., D. S., and T. L. acknowledge support from the Office of Naval Research MURI grant N00014-23-1-2567.


\clearpage
\onecolumngrid
\appendix

\begingroup
\footnotesize
\renewcommand{\arraystretch}{1.08}
\setlength{\tabcolsep}{2.25pt}
\setlength{\arraycolsep}{3.5pt}

\newcolumntype{C}[1]{>{\centering\arraybackslash}m{#1}}
\appendix*
\section{Symmetry-Constrained Tensor Forms}
\noindent
All point groups listed below are noncentrosymmetric and are given in the standard crystallographic setting associated with each SG. Here, “polar” means that the point group allows an invariant polar vector, corresponding to a symmetry-allowed polar axis, whereas “nonpolar” means that no nonzero polar vector is left invariant by all point group operations. Since all space groups considered here are noncentrosymmetric, the polar/nonpolar classification provides additional information beyond the absence of inversion symmetry: polar groups are necessarily noncentrosymmetric, but noncentrosymmetric groups need not be polar. In the point group properties column, ``proper'' means that all symmetry operations are proper rotations and therefore permit crystallographic chirality, whereas ``improper'' means that the point group contains at least one mirror or rotoinversion operation.

\medskip

\begin{table}[ht]
\caption{\label{tab:DK_shapes}Symmetry-constrained forms of $\mathbf{D}$ and $\mathbf{K}$ for all noncentrosymmetric space groups.}
\begin{center}
\begin{tabular}{
@{}
C{0.215\textwidth}
C{0.155\textwidth}
C{0.190\textwidth}
C{0.310\textwidth}
C{0.080\textwidth}
@{}
}
\hline\hline
\textbf{Noncentrosymmetric Space Group (SG)} &
\textbf{Point Group(s)} &
\textbf{Point Group Properties} &
$\mathbf{D}$ or $\mathbf{K}$ &
\textbf{\# SGs} \\
\hline\hline
\noalign{\vskip 3pt}

SG 1 &
$1$ &
polar, proper &
$\displaystyle
\left(
\begin{array}{ccc}
c_{xx} & c_{xy} & c_{xz} \\
c_{yx} & c_{yy} & c_{yz} \\
c_{zx} & c_{zy} & c_{zz}
\end{array}
\right)
$ &
1 \\[2pt]

SG 3--5 &
$2$ &
polar, proper &
$\displaystyle
\left(
\begin{array}{ccc}
c_{xx} & 0 & c_{xz} \\
0 & c_{yy} & 0 \\
c_{zx} & 0 & c_{zz}
\end{array}
\right)
$ &
3 \\[2pt]

SG 6--9 &
$\mathrm{m}$ &
polar, improper &
$\displaystyle
\left(
\begin{array}{ccc}
0 & c_{xy} & 0 \\
c_{yx} & 0 & c_{yz} \\
0 & c_{zy} & 0
\end{array}
\right)
$ &
4 \\[2pt]

SG 16--24 &
$222$ &
nonpolar, proper &
$\displaystyle
\left(
\begin{array}{ccc}
c_{xx} & 0 & 0 \\
0 & c_{yy} & 0 \\
0 & 0 & c_{zz}
\end{array}
\right)
$ &
9 \\[2pt]

SG 25--46 &
$\mathrm{mm}2$ &
polar, improper &
$\displaystyle
\left(
\begin{array}{ccc}
0 & c_{xy} & 0 \\
c_{yx} & 0 & 0 \\
0 & 0 & 0
\end{array}
\right)
$ &
22 \\[2pt]

SG 75--80, 143--146, 168--173 &
\shortstack{
$4$\\
$3$\\
$6$
} &
polar, proper &
$\displaystyle
\left(
\begin{array}{ccc}
c_{xx} & c_{xy} & 0 \\
-c_{xy} & c_{xx} & 0 \\
0 & 0 & c_{zz}
\end{array}
\right)
$ &
16 \\[2pt]

SG 81--82 &
$\overline{4}$ &
nonpolar, improper &
$\displaystyle
\left(
\begin{array}{ccc}
c_{xx} & c_{xy} & 0 \\
c_{xy} & -c_{xx} & 0 \\
0 & 0 & 0
\end{array}
\right)
$ &
2 \\[2pt]

SG 89--98, 149--155, 177--182 &
\shortstack{
$422$\\
$32$\\
$622$
} &
nonpolar, proper &
$\displaystyle
\left(
\begin{array}{ccc}
c_{xx} & 0 & 0 \\
0 & c_{xx} & 0 \\
0 & 0 & c_{zz}
\end{array}
\right)
$ &
23 \\[2pt]

SG 99--110, 156--161, 183--186 &
\shortstack{
$4\mathrm{mm}$\\
$3\mathrm{m}$\\
$6\mathrm{mm}$
} &
polar, improper &
$\displaystyle
\left(
\begin{array}{ccc}
0 & c_{xy} & 0 \\
-c_{xy} & 0 & 0 \\
0 & 0 & 0
\end{array}
\right)
$ &
22 \\[2pt]

SG 111--114, 121--122 &
$\overline{4}2\mathrm{m}$ &
nonpolar, improper &
$\displaystyle
\left(
\begin{array}{ccc}
c_{xx} & 0 & 0 \\
0 & -c_{xx} & 0 \\
0 & 0 & 0
\end{array}
\right)
$ &
6 \\[2pt]

SG 115--120 &
$\overline{4}\mathrm{m}2$ &
nonpolar, improper &
$\displaystyle
\left(
\begin{array}{ccc}
0 & c_{xy} & 0 \\
c_{xy} & 0 & 0 \\
0 & 0 & 0
\end{array}
\right)
$ &
6 \\[2pt]

SG 195--199, 207--214 &
\shortstack{
$23$\\
$432$
} &
nonpolar, proper &
$\displaystyle
\left(
\begin{array}{ccc}
c_{xx} & 0 & 0 \\
0 & c_{xx} & 0 \\
0 & 0 & c_{xx}
\end{array}
\right)
$ &
13 \\[2pt]

SG 174, 187--190, 215--220 &
\shortstack{
$\overline{6}$\\
$\overline{6}\mathrm{m}2$\\
$\overline{6}2\mathrm{m}$\\
$\overline{4}3\mathrm{m}$
} &
nonpolar, improper &
$\displaystyle
\left(
\begin{array}{ccc}
0 & 0 & 0 \\
0 & 0 & 0 \\
0 & 0 & 0
\end{array}
\right)
$ &
11 \\[2pt]

\noalign{\vskip 3pt}
\hline\hline
\end{tabular}
\end{center}
\end{table}

\newpage

\begin{table}[ht]
\caption{\label{tab:G_shapes}Symmetry-constrained forms of $\mathbf{G}$ for all noncentrosymmetric space groups.}
\begin{center}
\begin{tabular}{
@{}
C{0.215\textwidth}
C{0.155\textwidth}
C{0.190\textwidth}
C{0.310\textwidth}
C{0.080\textwidth}
@{}
}
\hline\hline
\textbf{Noncentrosymmetric Space Group (SG)} &
\textbf{Point Group(s)} &
\textbf{Point Group Properties} &
$\mathbf{G}$ &
\textbf{\# SGs} \\
\hline\hline
\noalign{\vskip 5pt}

SG 1 &
$1$ &
polar, proper &
$\displaystyle
\left(
\begin{array}{ccc}
c_{xx} & c_{xy} & c_{xz} \\
c_{yx} & c_{yy} & c_{yz} \\
c_{zx} & c_{zy} & c_{zz}
\end{array}
\right)
$ &
1 \\[7pt]

\multirow[c]{2}{*}{SG 3--9} &
\rule{0pt}{3.4ex}$2$ &
polar, proper &
\multirow[c]{2}{*}{$\displaystyle
\left(
\begin{array}{ccc}
c_{xx} & 0 & c_{xz} \\
0 & c_{yy} & 0 \\
c_{zx} & 0 & c_{zz}
\end{array}
\right)
$} &
\multirow[c]{2}{*}{7} \\
&
\rule{0pt}{3.4ex}$\mathrm{m}$ &
polar, improper &
&
\\[7pt]

\multirow[c]{2}{*}{SG 16--46} &
\rule{0pt}{3.4ex}$222$ &
nonpolar, proper &
\multirow[c]{2}{*}{$\displaystyle
\left(
\begin{array}{ccc}
c_{xx} & 0 & 0 \\
0 & c_{yy} & 0 \\
0 & 0 & c_{zz}
\end{array}
\right)
$} &
\multirow[c]{2}{*}{31} \\
&
\rule{0pt}{3.4ex}$\mathrm{mm}2$ &
polar, improper &
&
\\[7pt]

\multirow[c]{2}{*}{
\shortstack{
SG 75--82, 143--146,\\
168--174
}
} &
\rule{0pt}{3.4ex}$4,\ 3,\ 6$ &
polar, proper &
\multirow[c]{2}{*}{$\displaystyle
\left(
\begin{array}{ccc}
c_{xx} & c_{xy} & 0 \\
-c_{xy} & c_{xx} & 0 \\
0 & 0 & c_{zz}
\end{array}
\right)
$} &
\multirow[c]{2}{*}{19} \\
&
\rule{0pt}{3.4ex}$\overline{4},\ \overline{6}$ &
nonpolar, improper &
&
\\[7pt]

\multirow[c]{4}{*}{
\shortstack{
SG 89--122, 149--161,\\
177--190
}
} &
\rule{0pt}{3.4ex}$422,\ 32,\ 622$ &
nonpolar, proper &
\multirow[c]{4}{*}{$\displaystyle
\left(
\begin{array}{ccc}
c_{xx} & 0 & 0 \\
0 & c_{xx} & 0 \\
0 & 0 & c_{zz}
\end{array}
\right)
$} &
\multirow[c]{4}{*}{61} \\
&
\rule{0pt}{3.4ex}$4\mathrm{mm},\ 3\mathrm{m},\ 6\mathrm{mm}$ &
polar, improper &
&
\\
&
\rule{0pt}{3.4ex}$\overline{4}2\mathrm{m},\
\overline{4}\mathrm{m}2$ &
nonpolar, improper &
&
\\
&
\rule{0pt}{3.4ex}$\overline{6}\mathrm{m}2,\
\overline{6}2\mathrm{m}$ &
nonpolar, improper &
&
\\[7pt]

\multirow[c]{2}{*}{
\shortstack{
SG 195--199,\\
207--220
}
} &
\rule{0pt}{3.4ex}$23,\ 432$ &
nonpolar, proper &
\multirow[c]{2}{*}{$\displaystyle
\left(
\begin{array}{ccc}
c_{xx} & 0 & 0 \\
0 & c_{xx} & 0 \\
0 & 0 & c_{xx}
\end{array}
\right)
$} &
\multirow[c]{2}{*}{19} \\
&
\rule{0pt}{3.4ex}$\overline{4}3\mathrm{m}$ &
nonpolar, improper &
&
\\[7pt]

\noalign{\vskip 3pt}
\hline\hline
\end{tabular}
\end{center}
\end{table}

\endgroup


\bibliography{manuscript}
\end{document}